\documentclass[twocolumn]{revtex4-2}
\usepackage[dvipdfmx]{graphicx}
\usepackage{amsmath,amsbsy,amssymb}
\usepackage{bm}
\usepackage{mathrsfs}
\usepackage{ulem}
\usepackage{textgreek}
\usepackage{mathrsfs}
\usepackage{multirow}
\usepackage{physics}
\usepackage{comment}

\usepackage{color}

\newcommand{\diff}{\mathrm{d}}

\newcommand{\imu}{\mathrm{i}}
\newcommand{\epn}{\mathrm{e}}

\newcommand{\ua}{\uparrow}
\newcommand{\da}{\downarrow}
\newcommand{\dg}{\dagger}

\newcommand{\al}{\alpha}
\newcommand{\sg}{\sigma}

\newcommand{\dvec}[1]{\hspace{-1mm}\stackrel{\leftrightarrow}{#1}\hspace{-1mm}}

\begin{document}

\title{
Generalized Slater-Condon parameters for relativistic strongly correlated orbitals
}

\author{
Shintaro Hoshino
}

\affiliation{
Department of Physics, Saitama University, Sakura, Saitama 338-8570, Japan 
}

\date{\today}

\begin{abstract}

Relativistic correction to the Coulomb interaction is considered for strongly correlated electron orbitals.
The atomic representation of the Coulomb-Breit interaction and its physical origin are clarified, to generalize a concept of the Slater-Condon parameters.
In the derivation, it proves advantageous to employ the spherical representation of photons, which can be categorized into three types: scalar, electric, and magnetic photons. The electronic degrees of freedom can similarly be classified in terms of the angular momentum tensor or multipoles. Consequently, the interaction between electric multipoles is mediated by scalar photons, magnetic multipoles by magnetic photons, and magnetic toroidal multipoles by electric photons.
These parameters can be integrated with the multiorbital Hubbard model or the Anderson lattice and serve as the foundational framework for investigating relativistic strongly correlated electron systems.

\end{abstract}

\maketitle

{\it Introduction.---}
In strongly correlated electron systems (SCES), the charge, spin and orbital degrees of freedom of nearly localized electrons are strongly entangled through electronic interactions. 
This gives rise to a variety of intriguing phenomena, including heavy electrons, unconventional superconductivity, and exotic electronic orderings. A notable example of strong correlation effects is observed in uranium-based materials, where the nearly localized $f$ electrons play a crucial role in low-energy behaviors. A distinguishing characteristic of this class of materials is the occurrence of spin-triplet superconductors, potentially realized in compounds like UPt$_{3}$ \cite{Fisher89,Machida12}, UBe$_{13}$ \cite{Ott84,Shimizu19}, UGe$_2$, URhGe, UCoGe, and UTe$_2$ \cite{Saxena00, Aoki01,Huy07,Ran19,Aoki19,Aoki_review}. Another fascinating system is URu$_2$Si$_2$ \cite{Mydosh11}, which exhibits hidden ordered states and unconventional superconductivity.

To comprehend the distinct behaviors of uranium-based materials and the factors that set them apart from other substances, it is essential to consider the role of $f$ orbitals. Due to the large atomic number of uranium (92), the spin-orbit coupling, which is a relativistic correction, plays a significant role. Considering the localized nature of these orbitals where interaction effects become relevant, we are tempted to explore the potential impact of a relativistic correction to the Coulomb repulsive interaction. This correction is commonly referred to as the Coulomb-Breit interaction \cite{Bethe_book, Itoh65} and has been investigated in the field of quantum chemistry \cite{Liu20}. More recently, it has also been studied for atoms with large atomic numbers using the local density approximation (LDA) \cite{Naito20}. However, in condensed matter systems such as SCES, this correction has yet to be thoroughly explored.
Spin-orbit coupling is typically small but is now considered even for $d$ electron systems and organic conductors, which demonstrates the importance of relativistic corrections in any materials.

Given the significance of relativistic corrections and interaction effects in SCES, this study focuses on the investigation of the Coulomb-Breit interaction within localized electron orbitals.
To explore the physics involved, theoretical approach using the multiorbital Anderson (Kondo) lattice and/or multiorbital Hubbard model are employed as a fundamental framework. In constructing this model, it is necessary to derive the interaction tensor denoted as $U_{ijkl}$ for a strongly correlated orbital.
The conventional Coulomb interaction is typically described by the Slater-Condon parameters \cite{Slater_book,Condon_book,Kanamori63}. 
For example, only four parameters $\mathcal F^{0,2,4,6}$ are required for $f$ orbitals.
These parameter sets also find application in the LDA+$U$ or LDA+DMFT approaches, which are relevant to real materials \cite{Czyzyk94,Aichhorn09}. The purpose of this paper is to elucidate the generalized form of the Slater-Condon parameters specifically designed for the Coulomb-Breit interaction.

Since the interaction tensor with four spin-orbital indices has complicated structure, it is desirable to explore the physical origin of each term and to obtain compact representation.
In the following, we employ the multipole expansion method as described in Ref.~\cite{Iimura21} and 
 consider the electron-photon interaction, which provides a simpler approach compared to the method involving direct electron-electron interactions.

{\it Coulomb-Breit interaction.---}
In condensed matter physics, it is useful to work with the Coulomb gauge for radiation of photons, since only the physical Hilbert space needs to be considered \cite{Bjorken_book,Sucher80}.
The most general Hamiltonian for electrons is written as 
\begin{align}
    \mathscr H 
&= \mathscr  H_{{\rm D},0}
+ \frac 1 2 \int \diff \bm r \diff \bm r' \frac{\rho_{\rm D}(\bm r)\rho_{\rm D}(\bm r')}{|\bm r-\bm r'|}  
\nonumber \\
&\hspace{3mm}
+\sum_{\bm k} \sum_{\al=1}^2 \hbar\omega_{\bm k} a_{\bm k\al}^\dg a_{\bm  k\al}
- \frac{1}{c} \int \diff \bm r\, \bm j_{\rm D} (\bm r)\cdot \bm A (\bm r)
, \label{eq:general_ham}
\end{align}
where $\omega_{\bm k} = ck$ ($k=|\bm k|$) is the dispersion relation for photons with Planck constant $\hbar$ and speed of light $c$.
The suffix `${\rm D}$' indicates that the physical quantities is defined in terms of the four component Dirac field.
$\mathscr H_{\rm D,0}$ is a one-body electron part, which includes, for example, the Coulomb potential from positively-charged nucleus.
The Coulomb interaction between electrons appear in the second term of $\mathscr H$ and is written in terms of the charge density operator $\rho_{\rm D}(\bm r)$.
The third term represents the energy of transverse photons, where $a_{\bm k \al}$ is an annihilation operator of photons with wavevector $\bm k$ and polarization $\al = 1,2$.
The fourth term is a coupling between electronic current density operator $\bm j_{\rm D}(\bm r)$ and photons.
The vector potential is expanded by plane waves as \cite{Berestetskii_book,Bjorken_book}
\begin{align}
    \bm A(\bm r) &= \sqrt{\frac{4\pi}{V}}\, c\sum_{\bm k \al} \sqrt{\frac{\hbar}{2\omega_{\bm k}}} \,
\bm \epsilon_\al(\hat {\bm k}) \qty( \epn^{\imu \bm k\cdot \bm r} a_{\bm k\al} + \epn^{-\imu \bm k\cdot \bm r} a_{\bm k\al}^\dg )
, \label{eq:photon_expansion}
\end{align}
where $\hat {\bm k}$ is a unit vector along $\bm k$, $\bm \epsilon_\al$ is a linearly polarized vector satisfying $\hat {\bm k}\cdot \bm \epsilon_\al =  \bm \epsilon_1 \cdot  \bm \epsilon_2 =  0$, and $V = \int \diff \bm r 1$ is a volume of system.
This vector potential describes the {\it internal} electromagnetic fields degrees of freedom in an isolated physical system.
When the {\it external} field is considered, the additional contribution $\bm A_{\rm ext}
$ needs to be added \cite{Bjorken_book}.

Now we apply the concrete expression of the charge density and current density in the non-relativistic limit up to $O(c^{-2})$ \cite{Wang06,Hoshino23}:
\begin{align}
    \rho_{\rm D} &\simeq \qty( 1 + \frac{\lambdabar^2}{8}\bm \nabla^2)\rho - \bm \nabla \cdot \bm P_S
    , \label{eq:rho_NRL}
    \\
    \bm j_{\rm D} &\simeq \bm j + c \bm \nabla \times \bm M_S
    , \label{eq:j_NRL}
\end{align}
where $\psi = (\psi_\ua, \psi_\da)^{\rm T}$ is the two-component Schr\"odinger field.
Each microscopic quantity is explicitly given by
\begin{align}
    &\rho = e \psi^\dg \psi
    ,\ \ \ 
    \bm j = - \frac{\imu ce \lambdabar}{2} \psi^\dg \dvec{\bm \nabla} \psi
    ,
    \nonumber \\
    &\bm M_S = \frac{e \lambdabar}{2} \psi^\dg \bm \sg \psi
    ,\ \ \ 
    \bm P_S = - \frac{\imu e \lambdabar^2}{8} \psi^\dg \dvec{\bm \nabla} \times \bm \sg \psi
    , \label{eq:NRLs}
\end{align}
where $\psi^\dg \dvec{\partial} \psi = \psi^\dg \partial \psi - (\partial \psi^\dg)\psi$ and $\bm \sg$ is a Pauli matrix. 
$\lambdabar = \frac{\hbar}{mc}$ with electron mass $m$ is the reduced Compton length, which is a minimal length in the relativistic quantum mechanics of electrons \cite{Berestetskii_book}.

We can trace out the photon degrees of freedom without loss of generality and effective Hamiltonian for electrons is obtained.
In the non-relativistic limit, we obtain the following Coulomb-Breit interaction \cite{Itoh65}:
\begin{align}
    \mathscr H_{\rm C} &\simeq 
    \frac 1 2 \int \frac{\rho_1\rho_2}{r}
    -\frac{\pi \lambdabar^2}{2} \int \delta(\bm r)\rho_1\rho_2
    - \int \frac{\bm r \cdot \bm P_{S1} \, \rho_2}{r^3}
    \label{eq:Coulomb}
\end{align}
for the Coulomb part and
\begin{align}
    &\mathscr H_{\rm B} = 
    -\frac{1}{4c^2} \int 
    \frac{1}{r} \qty[\bm j_1\cdot \bm j_2 
    + \frac{(\bm j_1\cdot \bm r)(\bm j_2\cdot \bm r)}{r^2}]
    \nonumber \\
    &\ \ \ -\frac 1 c\int \frac{(\bm r\times \bm M_{S1}) \cdot \bm j_2}{r^3}
    + \frac 1 2 \int \bigg[-\frac{8\pi}{3} \delta(\bm r)\bm M_{S1}\cdot \bm M_{S2}
    \nonumber \\
    &\ \ \ \ \ \ 
    + \qty( \frac{ \bm M_{S1}\cdot\bm M_{S2} }{r^3} -\frac{3(\bm M_{S1}\cdot \bm r)(\bm M_{S2}\cdot \bm r)}{r^5})' \, \bigg]
    \label{eq:Breit}
\end{align}
for the Breit part mediated by transverse photons.
A simple derivation is provided in Supplementary Material (SM) \cite{suppl}.
We have used short-hand notations such as $\int \equiv \int\diff \bm r_1\diff \bm r_2$, $O_{1,2}\equiv O(\bm r_{1,2})$, and $\bm r = \bm r_1 - \bm r_2$.
The prime symbol ($'$) indicates that the integral is performed without contribution at $\bm r=\bm 0$ \cite{Bethe_book}.
Note that the Hamiltonian is described in a second quantization form and the normal ordering is implicitly assumed in the interaction term.

The first line of Eq.~\eqref{eq:Breit} represents the attraction between parallel currents and it can be a glue for local electron pairing in relativistic atoms.
Such attraction has been suggested to cause the Amperean pairing \cite{Khveshchenko93,Lee07,Lee14}.
Recently, the cavity-mediated superconducitivty has also been studied \cite{Schlawin19,Gao20,Chakraborty21,Schlawin22,Bloch22}, in which the photons mediate the attraction among electrons.
As shown below, we construct a foundation for exploring such exotic physics in SCES by clarifying the local representation of the Coulomb-Breit interaction, i.e., generalized Slater-Condon parameters.

In order to derive the atomic representation, however, the above electron-only representation is hard to be handled because of the singular behavior at $\bm r_1 \simeq \bm r_2$.
In contrast, this difficulty does not appear for the conventional Coulomb interaction, where the standard Laplace expansion for $|\bm r_1-\bm r_2|^{-1}$ can simply be used.
As shown below, it appears that it is more suitable to first deal with the atomic representation of the electron-photon coupling without tracing out photons.
Since the Coulomb interaction is written by only the electronic degrees of freedom in Eq.~\eqref{eq:general_ham}, we manipulate it with path-integral formalism to facilitate a unified depiction of the Coulomb-Breit interaction based on the coupling to photons.

{\it Introduction of scalar photon.---}
The partition function is given by the imaginary time action $\mathscr S = \int_0^\beta \diff \tau \mathscr L$ where $\mathscr L$ is the corresponding Lagrangian and $\beta$ is inverse temperature \cite{Negele_book}.
Applying the Stratonovich-Hubbard transformation for the Coulomb interaction part, we obtain 
\begin{align}
    &\mathscr L_{\rm eff} = 
    \sum_{\bm k}\sum_{\al=1}^2 a_{\bm k\al}^* \partial_\tau a_{\bm k\al} 
    + 
    \sum_{\bm k}\sum_{\al=0}^2 \hbar \omega_{\bm k} a_{\bm k\al}^*  a_{\bm k\al}
    \nonumber \\
    &- \sqrt{\frac{2\pi\hbar}{c V}} \int \diff \bm r \sum_{\bm k}\sum_{\al=0}^2 \frac{1}{\sqrt k}
    j_\al(\bm r;\hat{\bm k})
\qty( \epn^{\imu \bm k\cdot \bm r} a_{\bm k\al} + 
{\rm c.c.}) 
    , \label{eq:effective_lagrangian}
\end{align}
where the physical degrees of freedom are described by complex and Grassmann numbers.
Although one-body part for the Stratonovich-Hubbard field  usually appears with the factor $2\pi /|\bm k|^2$ \cite{Nagaosa_book}, we have employed another definition of the field.
We have defined the matter fields
\begin{align}
    j_0(\bm r;\hat{\bm k}) &= \imu c \rho_{\rm D}(\bm r)
    , \label{eq:source0}
    \\
    j_{1,2}(\bm r;\hat{\bm k}) &= \bm j_{\rm D}(\bm r) \cdot \bm \epsilon_{1,2} (\hat{\bm k})
    . \label{eq:source1}
\end{align}
With Eq.~\eqref{eq:effective_lagrangian}, it is clear that the Coulomb interaction is mediated by the scalar photon $a_{\bm k0}$.
The differences between scalar photon ($\al=0$) and transverse photons ($\al=1,2$) are 
(i) the presence or absence of the Berry phase term $\int a^* \partial_\tau a$ in the first term of Eq.~\eqref{eq:effective_lagrangian} and 
(ii) imaginary unit appears for the scalar photon case in Eq.~\eqref{eq:source0}, which results from repulsion between charges (Coulomb interaction).
On the other hand, the attractive interaction is obtained for currents mediated by transverse photons (Breit interaction).
In the non-relativistic limit, the retardation effect associated with the Berry phase term can be neglected in the leading-order approximation with respect to $1/c$ expansion \cite{Gorceix88}.
Thus, the scalar and transverse photons are described in a unified way in the non-relativistic limit relevant to condensed matter physics.

{\it Spherical-wave expansion of photons.---}
We now seek for the representation in the atomic wave function basis in order to derive the generalized Slater-Condon parameters.
We first introduce the spherical photon operators, which are defined from the transverse photons by \cite{Berestetskii_book, Messiah_book, Rose_book}
\begin{align}
    a_{JM}^{(\lambda)} (k) = \sqrt{\frac{V}{(2\pi)^3}}\ k \sum_{\al=1}^2 \int \diff \hat {\bm k} \ 
    \bm Y^{(\lambda)*}_{JM}(\hat{\bm k}) \cdot \bm \epsilon_\al(\hat{\bm k}) a_{\bm k\al}
    \label{eq:spherical_photon}
\end{align}
for $\lambda=1,0$, each of which corresponds to electric photon and magnetic photon, respectively.
This terminology is associated with the radiation from oscillating electric or magnetic dipoles \cite{Berestetskii_book}.
Here we consider the operator formalism, and the corresponding path-integral representation can then be easily sought.
The vector spherical harmonics $\bm Y^{(\lambda=-1,0,1)}_{JM}(\Omega)$ are given in terms of spherical harmonics $Y_{JM}(\Omega)$ as \cite{Varshalovich_book, Berestetskii_book}
\begin{align}
    \bm Y_{JM}^{(1)} &= [J(J+1)]^{-1/2} \ \bm \nabla_\Omega Y_{JM}
    , \label{eq:vec_harmonic_1}
    \\
    \bm Y_{JM}^{(0)} &= [J(J+1)]^{-1/2} \ 
    \bm n\times ( - \imu \bm \nabla_\Omega) Y_{JM}
    , \label{eq:vec_harmonic_0}
    \\
    \bm Y_{JM}^{(-1)} &= \bm n Y_{JM}
    , \label{eq:vec_harmonic_m1}
\end{align}
where $\bm r =(r,\Omega)$ is a spherical coordinate
with $\bm n = \bm r/r$.
The spatial derivative is given by $\bm \nabla = {\bm n}\frac{\partial}{\partial r} + \frac 1 r \bm \nabla_\Omega$.
Equation~\eqref{eq:spherical_photon} is interpreted as a basis transformation from the plane wave ($\bm k\al$) to the spherical wave ($kJM\lambda$).
The vector potential is then rewritten as
\begin{align}
\hspace{-3mm}
    \bm A(\bm r) = 
    \frac{\sqrt{\hbar c}}{2\pi} 
    \int \frac{\diff \bm k}{k^{3/2}}
    \sum_{\lambda=0}^1
    \sum_{JM}\qty(
    \epn^{\imu \bm k\cdot \bm r} \bm Y^{(\lambda)}_{JM}(\hat{\bm k}) a_{JM}^{(\lambda)}(k)
    +{\rm H.c.}
    )
\end{align}
which is a useful expression when combined with atomic wave functions of electrons.
A similar approach has been employed for the interactions between nuclear matter and electromagnetic field \cite{Rose_book}.
With the above definitions,
the commutation relation is given by
\begin{align}
\qty[ a_{JM}^{(\lambda)}(k), a_{J'M'}^{(\lambda')\dagger}(k')]
&= \delta(k-k') \delta_{\lambda\lambda'}\delta_{JJ'}\delta_{MM'}
\end{align}
for $\lambda=0,1$.

In the path-integral formalism, the above boson operators become a complex number.
In a similar manner, the scalar photon $a_{\bm k0}$ in Eq.~\eqref{eq:effective_lagrangian} is expanded by the spherical wave as
\begin{align}
    a^{(-1)}_{JM} (k) &= \sqrt{\frac{V}{(2\pi)^3}}\  k \int \diff \hat {\bm k} \   \bm n \cdot \bm Y_{JM}^{(-1)*} (\hat {\bm k}) a_{\bm k0}
    . \label{eq:scalar_spherical}
\end{align}
Noting the relation $\bm n \cdot \bm Y^{(-1)}_{JM} = Y_{JM}$, Eq.~\eqref{eq:scalar_spherical} is usual spherical harmonics expansion of a scalar function.
Hence, just for convenience, we assign the index $\lambda=-1$ for the scalar photons.
Note that longitudinal photon is absent in the Coulomb gauge condition.
We have summarized the properties of photons in Tab.~\ref{tab:photons}.

\begin{table}[t]
\caption{
List of photons in plane wave representation and in spherical representation with total angular momentum (or rank) $J$.
The signs from spatial inversion (SI) and time-reversal (TR) are also shown.
}
\begin{tabular}{c| c | c | c | c}
\hline
Basis Type & Index & Name & SI & TR
\\
\hline
plane wave&$\al=1,2$ & transverse photons & $-$ & $-$ \\
 $a_{\bm k\al}$&$\al=0$ & scalar photon & $+$ & $+$ \\
\hline
spherical wave&$\lambda=1$ & electric photon & $(-1)^J$ & $-$ \\
$a_{JM}^{(\lambda)} (k)$&$\lambda=0$ & magnetic photon & $(-1)^{J+1}$ & $-$ \\
&$\lambda=-1$ & scalar photon & $(-1)^J$ & $+$ \\
\hline
  \end{tabular}
\label{tab:photons}
\end{table}

The one-body part for photons in the Lagrangian is now written by
\begin{align}
    \mathscr L^0_{\rm photon} 
    &= \sum_{\lambda=-1}^1 \sum_{JM}\int_0^\infty \diff k \ \hbar c k\  a_{JM}^{(\lambda) *}(k) a^{(\lambda)}_{JM}(k)
    . \label{eq:spher_energy}
\end{align}
As already noted, the Berry connection part for electric and magnetic photons
is neglected 
in non-relativistic limit.
Thus the three (electric, magnetic, scalar) photons are handled in a unified manner also for the spherical representation.

{\it Multipole representation of electrons.---}
Next, we expand the electron field operator by the atomic orbitals as
\begin{align}
    \psi_\sg(\bm r) &= \sum_{nlm} R_{nl}(r) Y_{l m} (\Omega) c_{nlm\sg}
    ,
\end{align}
where $R$ is a radial wave function.
The integers $n,l,m$ are quantum numbers for principal, angular momentum and magnetic angular momentum, respectively ($m\in [-l,l]$).
Here we consider the localized orbitals of a hydrogen-like atom, and the spread of wave function is characterized by the Bohr radius $a = \frac{\hbar^2}{mZe^2}$ with effective nuclear charge $Z$.
In a typical SCES, we consider the correlation effect within a single angular momentum orbital such as $3d$ or $5f$ orbital.
Hence we focus on a fixed $(n,l)$ subspace in this paper.

The multipole representation of Coulomb interaction has been sought in solid state materials \cite{Sugano_book,Bunemann17,Iimura21,Letouze23}, and we extend this approach to the Coulomb-Breit interaction.
We employ the multipole or angular momentum tensor expansion of the electronic degrees of freedom.
We define the operator by \cite{Wang17}
\begin{align}
    T_{LS}^{JM} &= \sum_{mm'}\sum_{\sg\sg'} \qty( O^{JM}_{LS})_{m\sg,m'\sg'} c^\dg_{m\sg} c_{m'\sg'}
    ,
\end{align}
where the index $(n,l)$ in annihilation operator is omitted for simplicity.
The condition $|L-S| \leq J \leq L+S$ ($0\leq L\leq 2l$ and $S=0,1$) needs to be satisfied \cite{suppl}.
The above matrix is normalized by
${\Tr\,}
O_{LS}^{JM\dg} O_{LS}^{JM}= \sum_{m\sg}1 = 2(2l+1)$.
Because of the completeness of the angular momentum tensor, the inverse relation from $T$ to $c^\dg c$ is also obtained \cite{Iimura21}.

With the above representation in terms of spherical waves, 
the interaction between electrons and photons in Lagrangian is now represented as follows:
\begin{align}
    \mathscr L_{\rm int} &= - \sum_{\lambda=-1}^1 \sum_{J=0}^\infty \sum_{M=-J}^J
     \sum_{LS}
    \int_0^\infty \diff k \sqrt{\hbar c k}
    \nonumber \\
    &\ \ \ 
     \times \Big[
     g^{(\lambda)}_{JLS} (k)
     a_{JM}^{(\lambda)} (k) T_{LS}^{JM} + s_\lambda ({\rm c.c.}) \Big]
     , \label{eq:spher_coupling}
\end{align}
where the path-integral representation is employed (see SM for the derivation \cite{suppl}).
We have introduced the `sign function' $s_\lambda$: $s_{0,1} = 1$ and $s_{-1}=-1$.
Namely, the complex conjugated quantity is added for $\lambda=0,1$ and is subtracted for $\lambda=-1$.
We have also performed the spherical wave expansion of the plane wave $\epn^{\imu \bm k\cdot \bm r}$, and the integral with respect to radius $r$ is performed numerically.
Explicit forms of the coupling constant $g_{JLS}^{(\lambda)}(k)$ is provided in SM \cite{suppl}.

It is useful to point out the relation of the above quantity to electronic multipoles in condensed matter \cite{Ohkawa83,Shiina97,Kuramoto00,Santini00,Kiss05,Takimoto05,Kusunose08,Kuramoto09,Haule09,Ikeda12,Suzuki17,Hayami18,Chikano21,Hoshino23}.
In localized electron orbitals, the electronic operator is classified by the spatial inversion (SI), time-reversal (TR), and even-odd of the total angular momentum (rank, $J$).
For a fixed angular momentum $l$, only the even spatial parity $(-1)^{2l}=1$ is allowed.
Then we have the following four kinds of multipoles: electric (TR=$+$, even-rank), magnetic (TR=$-$, odd-rank), electric toroidal (TR=$+$, odd-rank), and magnetic toroidal (TR=$-$, even-rank) multipoles \cite{Kusunose20,Hoshino23}.
On the other hand, the photon operators are classified as electric, magnetic, and scalar photons.
The magnetic photon has a spatial parity $(-1)^{J+1}$, and electric photon has $(-1)^{J}$ \cite{Berestetskii_book}.
The same angular momentum $(J,M)$ for electron and photon operators is shared due to the angular momentum conservation.
Because of TR$=-$ for transverse photons, 
for a fixed-$l$ subspace, the magnetic odd-rank photon couples to magnetic multipoles, and electric even-rank photon to magnetic toroidal multipoles.
In addition, the scalar photon has SI$=(-1)^J$ with TR$=+$, and then the scalar even-rank photon couples to electric multipoles, and electric toroidal multipole is absent if we consider a fixed $l$.

Now we are ready to trace out photons.
The photon part of action is composed of Eqs.~\eqref{eq:spher_energy} and \eqref{eq:spher_coupling}, and the photon field can be integrated out.
Then we obtain the effective interaction among electrons
\begin{align}
    \mathscr L_{\rm eff, int}
     &= \frac 1 2 \sum_{\lambda=-1}^1 \sum_{JM} 
     \sum_{L_1S_1L_2S_2}
     I^{J(\lambda)}_{L_1S_1, L_2S_2} 
     \ T_{L_1S_1}^{JM*} 
     \ T_{L_2S_2}^{JM}
     ,
\end{align}
where only the contributions up to $O(c^{-2})$ are kept.
The total angular momentum $J$ (or rank) is a good quantum number in a spherical system, while different $(L,S)$ are mixed in general.
In our previous work \cite{Iimura21}, it has been pointed out that the odd-rank component is absent in the conventional Coulomb interaction even when the screening effect is considered in solids.
The reason for this is now clarified: the odd-rank contribution originates from the Breit interactions mediated by magnetic photons.

{\it Generalized Slater-Condon parameters.---}
The interaction parameter $I$ can be explicitly evaluated by using the atomic wave function, and is written as
\begin{align}
    I_{L_1S_1,L_2S_2}^{J(\lambda)}
    &= F_{L_1S_1,L_2S_2}^{J(\lambda)}
    + 
    \qty(\frac{\lambdabar}{a})^2 G_{L_1S_1,L_2S_2}^{J(\lambda)}
    . \label{eq:GSCparam}
\end{align}
The first term is a conventional Coulomb interaction, and the parameter is finite when $S_1=S_2=0$, $L_1=L_2=J$ for even $J$ and $\lambda=-1$, which indicates that the interaction is mediated by scalar photon.
On the other hand, the second term with $G$ shows a relativistic correction, where
the factor $\lambdabar/a$ is identical to $Z\al<1$ with $\al = \frac{e^2}{\hbar c} \simeq 1/137$.
Since $F$ is related to conventional Slater-Condon parameters $\mathcal F^{0,2,4,\cdots}$ \cite{suppl},  Eq.~\eqref{eq:GSCparam} is regarded as their generalized version.

\begin{figure}[tb]
    \centering
    \includegraphics[width=90mm]{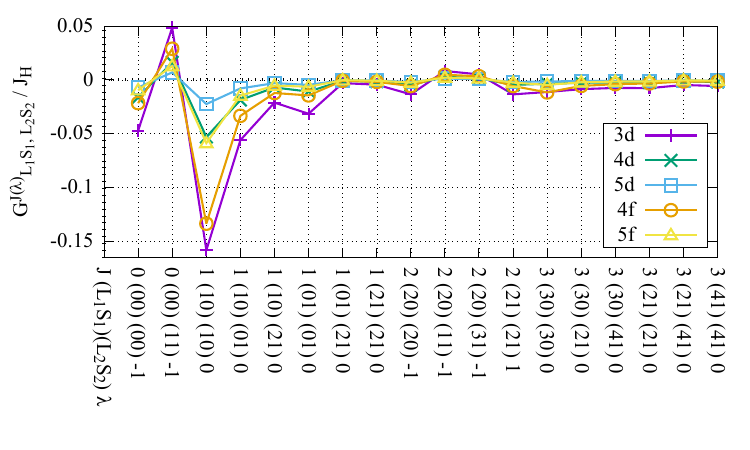}
    \caption{
    List of the coefficients in interaction parameter $G_{L_1S_1,L_2S_2}^{J(\lambda)}$ ($=G_{L_2S_2,L_1S_1}^{J(\lambda)}$) up to rank $J=3$.
The values are normalized by the bare Hund's coupling $J_{\rm H}$ in the atomic limit for each $n,l$.
The interactions are mediated by the type-$\lambda$ photons.
}
\label{fig:values}
\end{figure}

Instead of showing the bare numerical value of $I$, 
we show the values relative to well-known parameter.
Since the bare Coulomb interaction of the rank zero, i.e. the Slater-Condon parameter $\mathcal F^0$, is strongly screened in a lattice environment, we choose the Hund's coupling $J_{\rm H}$, which is related to the higher-order Slater-Condon parameters $\mathcal F^{2,4,\cdots}$ \cite{Marel88,suppl}.
Figure~\ref{fig:values} summarizes the numerical values of $G_{L_1S_1,L_2S_2}^{J\lambda} / J_{\rm H}$ up to rank $J=3$ for $l = 2$ ($d$-orbital) and $l=3$ ($f$-orbital).
The highest rank is $J=2l+1$ (not shown), and the values become basically smaller when the rank becomes higher.
For comparison, the conventional Coulomb interaction parameter $F^{J(\lambda)}_{L_1S_1,L_2S_2}$ relative to the Hund's coupling is also listed in SM \cite{suppl}.

Let us first take a look at $\lambda=-1$ (scalar photon) contributions.
An attractive interaction correction $G^{J(-1)}_{00,00}$ for $J=0,2$ enters through the $\lambdabar^2\bm \nabla^2 \rho$ term in Eq.~\eqref{eq:rho_NRL}.
The other terms with $J=2$ originates from the coupling between charge $\rho_0$ and polarization $\bm P_S$ in Eq.~\eqref{eq:Coulomb}, since the different spins $S_1=0$ are $S_2=1$ are involved.

Next we consider the $\lambda=0$ sector (magnetic photon), which is finite only when $J$ is odd.
The largest contribution enters through $L_1=L_2=1$ and $S_1=S_2=0$.
This contribution derives from current-current (or momentum-momentum) interaction in Eq.~\eqref{eq:Breit} since the spin is not directly involved.
The current-spin interaction also exists as e.g. $G^{1(0)}_{11,01}$, whose parameter is smaller than the current-current case.
The spin-spin type $G^{1(0)}_{01,01}$ has further smaller values.
On the other hand, the $\lambda=1$ (electric photon) contribution enters only for even $J$, whose value is smaller than the ones mediated by magnetic photons.

The present results show a compact and physically transparent representation of relativistic correction to the interaction and should be a foundation for the further exploration of a variety of correlated phenomena in SCES.

{\it Discussion and Summary.---}
Let us estimate the magnitude of the parameters for the Coulomb-Breit interaction in atomic limit.
As shown in Eq.~\eqref{eq:GSCparam}, the interaction parameter is written as $I \propto (Z\al)^2 J_{\rm H}$ where the proportionality constant is shown in Fig.~\ref{fig:values}.
In actual materials, the screening effects enter and the conventional Slater-Condon parameters becomes smaller compared to the bare values.
The screening is strongest for the isotropic component $\mathcal F^0$ and is much weaker for the anisotropic part $\mathcal F^{2,4,6}$.
Hence we expect that the relativistic corrections have further small screening effects.

For the concrete estimation, the effective nuclear charge $Z$ is needed.
Using the energy splitting in experimental data \cite{AtomicWebSite} and comparing it with atomic spin-orbit coupling, we obtain $Z=23$ for Ce$^{3+}$ [$(4f)^1$] and $Z=34$ for Yb$^{3+}$ [$(4f)^{13}$], which are consistent with the results in Ref.~\cite{Clementi67}.
As for $5f$ electrons, since no data is available for U atom, we take the excited states of Th$^{+}$ with the configuration $(5f)^1 (7s)^2$ \cite{AtomicWebSite}, which gives $Z=32$.
Assuming that there are no screening effects and using the bare value of $J_{\rm H}$, the largest relativistic correction part $I^{1(0)}_{10,10}$ is estimated as $65$K for Ce, $210$K for Yb, and  $40$K for Th.
Considering that the relativistic corrections include many terms with different $J,L_1S_1, L_2S_2$, we expect that the resulting energy scale is a few hundreds of K.
We note that the actual energy scale enters depending on the number of electrons since we consider the interaction effects.

Such a small energy scale can be relevant if the crystalline field splitting is small as in nearly localized $f$ electron systems.
Also, the heavy Fermi liquid is formed in some $f$-electron systems, where the effective band width is renormalized to be hundreds times smaller. 
Hence, the small but residual Coulomb-Breit interaction is expected to affect the low-energy physical behavior of heavy electrons.
The resulting physics, potentially encompassing phenomena like photon-mediated superconductivity, awaits investigation in $f$ electron systems.

For more quantitative description, 
the $f$ orbital may not be sufficient, and the other orbitals such as $d$-orbital should also be involved.
The corresponding interaction parameters can be calculated by using the technique presented in this paper.
In addition, the effective nuclear potential for many-electron atoms is different from the hydrogen-like $Z|e|/r$ potential.
These complications remain to be explored in the future.

To summarize, 
we have developed the generalized Slater-Condon parameters for the Coulomb-Breit interaction.
The physical consequence of the interaction remains to be clarified, which is expected to shed more light on the mysteries of electronic ordering in $f$ electron systems.

\section*{Acknowledgement}
The author is grateful to H. Ikeda, M.-T. Suzuki, M. Hirayama, and T. Naito for useful discussions.
This work was supported by KAKENHI Grants 
No.~21K03459 
and
No.~23H01130. 

\clearpage

\makeatletter
\renewcommand{\thepage}{S\arabic{page}}
\renewcommand{\thesection}{S\arabic{section}}
\renewcommand{\theequation}{S\arabic{equation}}
\renewcommand{\thefigure}{S\arabic{figure}}
\renewcommand{\thetable}{S\arabic{table}}
\makeatother

\setcounter{page}{1}
\setcounter{section}{0}
\setcounter{equation}{0}
\setcounter{table}{0}
\setcounter{figure}{0}

\noindent
{\bf SUPPLEMENTARY MATERIAL 
FOR \\
``
Generalized Slater-Condon parameters for \\
relativistic strongly correlated orbitals
''
}
\\[2mm]
Shintaro Hoshino
\\[2mm]
(Dated: \today)

\section{Simple derivation of Coulomb-Breit interaction}

While the Coulomb-Breit interaction is explicitly derived in Ref.~\cite{Itoh65}, here we derive it using second quantization form in a simpler manner.
We begin with the general Hamiltonian [Eq.~\eqref{eq:general_ham} of the main text] in Coulomb gauge.
With the Fourier expansion form in Eq.~\eqref{eq:photon_expansion}, we rewrite the Hamiltonian involving photons as
\begin{align}
    \mathscr H' &= \sum_{\bm k} \sum_{\al=1}^2 \hbar\omega_{\bm k} a_{\bm k\al}^\dg a_{\bm  k\al}
- \frac{1}{c} \int \diff \bm r\, \bm j_{\rm D} (\bm r)\cdot \bm A (\bm r)
    \\
    &= \sum_{\bm k\al}
    \hbar\omega_{\bm k} \tilde a_{\bm k\al}^\dg \tilde a_{\bm  k\al}
    - \sum_{\bm k\al} \frac{F_{\bm k\al}^\dg F_{\bm k\al}} {\hbar \omega_{\bm k}}
    , \label{eq:derive_Breit1}
\end{align}
where we have defined `shifted' photon field
\begin{align}
    \tilde a_{\bm k\al} &= a_{\bm k\al} - F_{\bm k\al}
    , \\
    F_{\bm k\al} &= \sqrt{\frac{2\pi \hbar}{\omega_{\bm k}}}\int \diff \bm r \bm j_{\rm D}(\bm r) \cdot \bm \epsilon_\al (\hat {\bm k}) \epn^{\imu \bm k\cdot \bm r}
    .
\end{align}
Assuming that the shifted photon parts are effectively independent of the electronic parts, we now concentrate on the second term of Eq.~\eqref{eq:derive_Breit1}.
The wave vector integral can be performed to give the Breit interaction:
\begin{align}
    \mathscr H_{\rm B} &= - \frac{1}{4c^2} \int \diff \bm r_1 \diff \bm r_2 \qty[ \frac{\bm j_{\rm D1}\cdot \bm j_{\rm D2}}{r} + \frac{(\bm j_{\rm D1}\cdot \bm r) (\bm j_{\rm D2}\cdot \bm r) }{r^3}]
    \label{eq:Breit_Dirac}
    ,
\end{align}
where $\bm j_{\rm D1,2}=\bm j_{\rm D}(\bm r_{1,2})$ and $\bm r=\bm r_1 - \bm r_2$.

Note that the more rigorous derivation should be done by employing the path-integral formalism, although the above trick provides a simple derivation.
Namely, the integration of photon fields can be explicitly performed, and then the static limit of the retarded interaction is taken to give the same expression derived above.
The retarded part is higher-order in $1/c$ expansion and is neglected.

Now we seek for the expression in non-relativistic limit.
Substituting Eqs.~\eqref{eq:rho_NRL} and \eqref{eq:j_NRL} into the Coulomb interaction in Eq.~\eqref{eq:general_ham} and the Breit interaction in Eq.~\eqref{eq:Breit_Dirac}, and using the following vector formula
\begin{align}
    \partial_i \partial_j \frac{1}{r} = - \frac{4\pi}{3} \delta_{ij}\delta(\bm r) + \qty( \frac{3x_ix_j-r^2\delta_{ij}}{r^5} )'
    ,
\end{align}
we arrive at Eq.~\eqref{eq:Breit} of the main text.
The resultant expression is identical to the second-quantized version of Ref.~\cite{Itoh65}.
Note also that the obtained expression implicitly assumes the normal ordering, which is naturally taken into account in the path-integral formalism.

\section{Atomic wave function}

We expand the field operator for each strongly correlated electron orbitals with principal and angular momentum quantum numbers $(n,l)$ as follows:
\begin{align}
    \psi_\sg(\bm r) &= \sum_{ m} R(r) Y_{l m} (\Omega) c_{m\sg}
    , \\
    R(r)
    &= b^{-3/2} \sqrt{\frac{(n-l-1)!}{2n(n+l)!}}
    \ \epn^{- \rho/2} \rho^\ell L_{n-l-1}^{2l + 1}(\rho)
    , \label{eq:atomic_radial}
\end{align}
where $\rho = r/b $ and $ b =na/2$ with the Bohr radius $a = \frac{\hbar^2}{mZe^2}$.
$R(r)$ is the radial wave function depending on $(n,l)$, $Y(\Omega)$ is the spherical harmonics, and $L(\rho)$ is the Laguerre polynomials.

\section{Angular momentum tensor}
We introduce the angular momentum tensor for a fixed angular momentum quantum number $l$ \cite{Varshalovich_book}:
\begin{align}
    (O_{LL_z})_{mm'} &= (-1)^{2L} C^{lm}_{lm'LL_z}
    , \\
    (O_{SS_z})_{\sg\sg'} &= (-1)^{2S} C^{s\sg'}_{s\sg SS_z}
\end{align}
with $s=1/2$, $L=0,\cdots,2l$ and $S=0,1$.
The constant $C$ is the Clebsch-Gordan coefficient. 
Then we define \cite{Wang17}
\begin{align}
    (O_{LS}^{JM})_{m\sg,m'\sg'} &= A_{LS}^{JM}\sum_{L_zS_z}C^{JM}_{LL_zSS_z} (O_{LM})_{mm'} (O_{SS_z})_{\sg\sg'}
    .
\end{align}
The normalization constant $A$ is determined by the condition
${\Tr\,}
O_{LS}^{JM\dg} O_{LS}^{JM}={\Tr\,}1=2(2l+1)$.

The angular momentum tensor operator for electrons is written as
\begin{align}
T_{LS}^{JM}  &= \sum_{m\sg m'\sg'} (O^{JM}_{LS})_{m\sg,m'\sg'} 
 c^\dg_{m\sg} c_{m'\sg'} .
\end{align}
Specifically, $T_{00}^{00} = \sum_{m\sg} c^\dg_{m\sg}c_{m\sg}$ represents the number operator, and the inner product of angular and spin momenta is given by $\bm \ell \cdot \bm s = \sqrt{\frac{l(l+1)}{4}} T_{11}^{00}$.

\section{Spherical coordinates}

The covariant basis vectors are defined by \cite{Varshalovich_book,Rose_book}
\begin{align}
    \bm e_{+1} &= - \frac{1}{\sqrt 2} (\hat {\bm x} + \imu \hat {\bm y})
    , \\
    \bm e_{0} &= \hat {\bm z} 
    , \\
    \bm e_{-1} &=  \frac{1}{\sqrt 2} (\hat {\bm x} - \imu \hat {\bm y})
    ,
\end{align}
where the hat ($\hat\ $) symbol in the right-hand sides represents a unit vector.
The contravariant basis vectors are given by
\begin{align}
\bm e^\mu &= \bm e_\mu^* = (-1)^\mu \bm e_{-\mu}
.
\end{align}
Any vector (operator) $\bm A$ can be expanded as
\begin{align}
    \bm A &= \sum_\mu A^\mu \bm e_\mu = \sum_\mu A_\mu \bm e^\mu
    ,\\
    A_{\mu} &= \bm e_\mu\cdot \bm A
    , \ \ \ 
    A^\mu = \bm e^\mu \cdot \bm A 
    .
\end{align}
The vector $\bm A$ can be chosen as spatial coordinate $\bm r = (x,y,z)$, derivative $\bm \nabla = \partial /\partial \bm r$, or Pauli matrix $\bm  \sg = (\sg^x,\sg^y,\sg^z)$.

We have introduced the vector spherical harmonics in the main text.
They satisfy the orthogonal relation
\begin{align}
    \int \diff \Omega \, \bm Y^{(\lambda')*}_{J'M'} (\Omega) \cdot \bm Y^{(\lambda)}_{JM} (\Omega) = \delta_{J'J} \delta_{\lambda'\lambda} \delta_{M'M}
    ,
\end{align}
and the completeness
\begin{align}
    \sum_{\lambda JM} \qty[ \bm Y_{JM}^{(\lambda)}(\Omega')]_i \qty[\bm Y_{JM}^{(\lambda)}(\Omega)]^*_j = \delta_{ij} \delta(\Omega - \Omega')
\end{align}
where $i,j = x,y,z$.
The contravariant component is written as
\begin{align}
    Y^{(\lambda)\mu}_{JM}(\Omega) &= \sum_{s=-1}^1 \mathcal Y_{JM}^{(\lambda)\mu}(s) Y_{J-s,M-\mu} (\Omega)
\end{align}
for $\mu=-1,0,1$,
which is useful in actual evaluations.
The coefficient $\mathcal Y$ is listed in Ref.~\cite{Varshalovich_book}.

\begin{widetext}

\section{Derivation of Generalized Slater-Condon parameters}

Here we derive the explicit form the spherical expansion for the coupling between matter and photon fields.
We begin with the spherical wave expansion of the vector potential:
\begin{align}
    \bm A(\bm r) &= \frac{\sqrt{2\pi \hbar c}}{(2\pi)^{3/2}} \int \frac{\diff \bm k}{k^{3/2}}
    \sum_{\lambda=0}^1
    \sum_{JM}\qty(
    \epn^{\imu \bm k\cdot \bm r} \bm Y^{(\lambda)}_{JM}(\hat{\bm k}) a_{JM}^{(\lambda)}(k)
    +{\rm c.c.}
    )
    .
\end{align}
We also consider the rotation
\begin{align}
    \bm \nabla \times \bm A &= 
    \frac{\sqrt{2\pi \hbar c}}{(2\pi)^{3/2}} \int \, \frac{\diff \bm k}{k^{1/2}}
    \sum_{\lambda =0}^1 \sum_{JM}  \qty(
    - \epn^{\imu \bm k\cdot \bm r} \bm Y^{(1-\lambda)}_{JM}(\hat{\bm k}) a_{JM}^{(\lambda)} (k)
    +{\rm c.c.})
    .
\end{align}
where we have used $\hat {\bm k}\times \bm Y_{JM}^{(\lambda)}(\hat {\bm k}) = \imu \bm Y_{JM}^{(1-\lambda)}(\hat {\bm k})$ for ($\lambda=0,1$) \cite{Varshalovich_book}.
In a similar manner, we expand the scalar potential defined by
\begin{align}
    \Phi(\bm r) 
    &\equiv
    \sqrt{\frac{2\pi V\hbar}{c}} \int \frac{\diff \bm k}{(2\pi)^3}\, \frac{1}{k^{1/2}}
    \qty(
    \epn^{\imu \bm k\cdot \bm r} a_{\bm k0}
    +{\rm c.c.}
    )
    \\
    &=
    \sqrt{\frac{2\pi \hbar}{c} } \int \frac{\diff \bm k}{(2\pi)^{3/2}}\, \frac{1}{k^{3/2}}
    \sum_{JM}\qty(
    \epn^{\imu \bm k\cdot \bm r} Y_{JM}(\hat{\bm k}) a_{JM}^{(-1)}(k)
    +{\rm c.c.}
    )
    ,
\end{align}
which originates from the Stratonovich-Hubbard field $a_{\bm k0}$.
We also consider its gradient
\begin{align}
    \bm \nabla\Phi(\bm r) 
    &= \sqrt{\frac{2\pi \hbar}{c} } \int \frac{\diff \bm k}{(2\pi)^{3/2}}\, \frac{1}{k^{3/2}}
    \sum_{JM}\qty(\imu k 
    \epn^{\imu \bm k\cdot \bm r} \bm Y^{(-1)}_{JM}(\hat{\bm k}) a^{(-1)}_{JM}(k)
    +{\rm c.c.}
    )
    .
\end{align}
The interaction Hamiltonian is then written as
\begin{align}
\mathscr H_{\rm int} &=
- \int \diff \bm r \qty( \frac 1 c \bm j_{\rm D}\cdot \bm A + \imu \rho_{\rm D} \Phi )
\\
&=
-
\int \diff \bm r \qty[ \frac 1 c
{\bm j} \cdot \bm A + \bm M_S \cdot (\bm \nabla \times \bm A)
+\imu \rho \qty(1+\frac{\lambdabar^2}{8}\bm \nabla^2) \Phi + \imu \bm P_S \cdot \bm \nabla \Phi
]
\\
&= -\frac{\sqrt{\hbar c}}{2\pi} \sum_{\lambda=0}^1\sum_{j=0}^\infty\sum_{M=-J}^J \int_0^\infty \diff k\sqrt k 
\qty[\mathcal J_{jm}^{(\lambda) } (k) + \mathcal M^{(\lambda) }_{JM}(k)]a_{jm}^{(\lambda)} (k) + {\rm c.c.}
\nonumber \\
&\ \ \ - \imu \frac{\sqrt{\hbar c}}{2\pi} \sum_{J=0}^\infty\sum_{M=-J}^J\int_0^\infty \diff k\sqrt k \qty[\rho_{JM} (k) + \mathcal P_{JM}(k)]a^{(-1)}_{JM}(k) - {\rm c.c.}
\label{eq:epcouple_spherical}
\end{align}
where
\begin{align}
    \mathcal J_{JM}^{(\lambda) } &= 
    \frac{2\pi e\lambdabar}{b} \sum_{s=-1}^1 Q_{J-s}(kb) \imu^{J-s} \sum_{\mu=-1}^1\sum_{m\sg m'\sg'}
    (\bm H_{M-\mu,mm'}^{j-s,ll})_\mu \delta_{\sg\sg'} 
    \mathcal Y^{(\lambda) \mu}_{JM}(s) c_{m\sg}^\dg c_{m'\sg'}
    \\
    &\equiv \frac{2\pi e\lambdabar}{b} \sum_{s=-1}^1 Q_{J-s}(\kappa) \sum_{LS} C^{(a)}_{J\lambda}(LS,s) T_{LS}^{JM}
    ,\\
\mathcal M_{JM}^{(\lambda) } &= 
    2\pi e\lambdabar \sum_{s=-1}^1 k P_{J-s}(kb) (-1) \imu^{J-s} \sum_{\mu=-1}^1\sum_{m\sg m'\sg'}
    G_{M-\mu,mm'}^{J-s,ll} (\bm \sg_{\sg\sg'})_\mu 
    \mathcal Y^{(1-\lambda) \mu}_{JM}(s)    
    \\
    &\equiv \frac{2\pi e\lambdabar}{b}\sum_{s=-1}^1 \kappa P_{J-s}(\kappa) \sum_{LS} C^{(b)}_{J\lambda}(LS,s) T_{LS}^{JM}
     ,   \\
\rho_{JM} &= 
    4\pi e \qty(1-\frac{\lambdabar^2}{8}k^2)P_{J}(kb) \imu^{J} \sum_{m\sg m'\sg'}
    G_{Mmm'}^{Jll} \delta_{\sg\sg'} 
     c_{m\sg}^\dg c_{m'\sg'}
    \\
    &\equiv 4\pi e \qty( 1-\frac{\lambdabar^2}{8b^2}\kappa^2) P_{J}(\kappa) \sum_{LS} C^{(c)}_{J}(LS) T_{LS}^{JM}
    ,    \\
 \mathcal P_{JM} &= 
     \frac{\pi e\lambdabar^2}{2b} \sum_{s=-1}^1 k Q_{J-s}(kb) \imu^{J-s+1} \sum_{\mu=-1}^1 \sum_{m\sg m'\sg'}
    (\bm H_{M-\mu,mm'}^{J-s,ll}\times \bm \sg_{\sg\sg'})_\mu 
    \mathcal Y^{(-1) \mu}_{JM}(s)  c_{m\sg}^\dg c_{m'\sg'}
    \\
    &\equiv \frac{\pi e \lambdabar^2}{2b^2} \sum_{s=-1}^1 \kappa Q_{J-s}(\kappa) \sum_{LS} C^{(d)}_{J}(LS,s) T_{LS}^{JM}
\end{align}
with $\kappa = kb$.
We have defined
\begin{align}
    G_{Mmm'}^{Lll} &=  \int \diff \Omega \ Y_{LM}(\Omega)\  Y^*_{lm}(\Omega) Y_{lm'}(\Omega)
    ,\ \ \ \ 
    \bm H _{Mmm'}^{Lll} = \int \diff \Omega \ Y_{LM}(\Omega) \  \qty[ Y^*_{lm}(\Omega)
    (-\imu \dvec{\bm \nabla}_\Omega)
    Y_{lm'}(\Omega) ]
\end{align}
for angular part and
\begin{align}
    P_L(\kappa) &=  \int_0^\infty \diff r \ r^2 R(r)^2 j_L(kr)
    ,\ \ \ \ 
    Q_L(\kappa) =  b \int_0^\infty \diff r \ r^1 R(r)^2 j_L(kr)
\end{align}
for radial part, which are dimensionless functions depending on $(n,l)$.
The contravariant components of the vector spherical harmonics $Y^{(\lambda)\mu}$ are defined in the last section.
$j_L(x)$ is the spherical Bessel function that appears by the Rayleigh expansion of $\epn^{\imu \bm k\cdot \bm r}$.
Note that the summation with respect to $LS$ is implicitly dependent on $J$.
Eq.~\eqref{eq:epcouple_spherical} coincides with Eq.~\eqref{eq:spher_coupling} in the main text.

The photon part and electron-photon interaction in the Lagrangian are now written as
\begin{align}
    \mathscr L' &= \sum_{\lambda=-1}^1 \sum_{JM} \int_0^\infty \diff k\qty[
    \hbar c k \  a_{JM}^{(\lambda) *} (k) a_{JM}^{(\lambda)} (k)
    -
    \sqrt {\hbar c k} \sum_{LS} 
    \qty( 
    g^{(\lambda)}_{JLS}(k) T_{LS}^{JM} a_{JM}^{(\lambda)} (k)
    + s_\lambda
    g^{(\lambda)*}_{JLS} T_{LS}^{JM*} a_{jm}^{(\lambda) *} (k) )
    ]   
    .
\end{align}
The Berry phase term is neglected in the leading-order approximation with respect to $1/c$ expansion.
The sign function is defined by $s_{0,1}=1$ and $s_{-1}=-1$, and
\begin{align}
    g^{(\lambda)}_{JLS} &= - \frac{e\lambdabar}{b} \sum_{s=-1}^1 \qty( Q_{J-s}(\kappa) C_{J\lambda}^{(a)}(LS,s) + \kappa P_{J-s}(\kappa) C^{(b)}_{J\lambda}(LS,s) )
\end{align}
for $\lambda=0,1$ and
\begin{align}
    g^{(\lambda=-1)}_{JLS}
    &=
    2\imu e P_{J}(\kappa) C_J^{(c)}(LS)
    +
    \frac{\imu  e\lambdabar^2}{4b^2} \qty(
    -\kappa^2 P_J(\kappa) C_J^{(c)}(LS)
    + \sum_{s=-1}^1 \kappa Q_{J-s}(\kappa) C_J^{(d)}(LS,s)
    )
    .
\end{align}
By tracing out spherical photons, the effective Lagrangian is given by
\begin{align}
    \mathscr L_{\rm eff} &= \frac 1 2 \sum_{J}\sum_{L_1S_1L_2S_2} 
    I^{J(\lambda)}_{L_1S_1,L_2S_2}
     T_{L_1S_1}^{JM} T_{L_2S_2}^{JM*}
     .
\end{align}
This interaction has a rotationally symmetric form.
The interaction parameter is defined by
\begin{align}
I^{J(\lambda)}_{L_1S_1,L_2S_2} &= - 2 s_\lambda \int_0^\infty \diff k \ 
     g^{(\lambda)}_{JL_1S_1}
     \ g^{(\lambda)*}_{JL_2S_2}
     .
\end{align}
This provides a generalized version of Slater-Condon parameters for Coulomb-Breit interaction.
For the evaluation of their numerical values, we write down
the detailed expressions as 
\begin{align}
    I^{J(\lambda)}_{L_1S_1,L_2S_2} &=
    \frac{2e^2\lambdabar^2}{b^3} \sum_{s_1s_2}\Big[
     R_{J-s_1,J-s_2}^{(a)} C^{(a)}_{J\lambda}(L_1S_1,s_1) C^{(a)}_{J\lambda}(L_2S_2,s_2) 
    +R_{J-s_1,J-s_2}^{(c)} C^{(b)}_{J\lambda}(L_1S_1,s_1) C^{(b)}_{J\lambda}(L_2S_2,s_2)
    \nonumber \\
    &\hspace{20mm}
    +R_{J-s_1,J-s_2}^{(b)} C^{(a)}_{J\lambda}(L_1S_1,s_1) C^{(b)}_{J\lambda}(L_2S_2,s_2) 
    +R_{J-s_2,J-s_1}^{(b)} C^{(b)}_{J\lambda}(L_1S_1,s_1) C^{(a)}_{J\lambda}(L_2S_2,s_2)
    \Big]
\end{align}
for $\lambda=0,1$, and
\begin{align}
    I^{J(\lambda=-1)}_{L_1S_1,L_2S_2} &= 
    \frac{8e^2}{b} R_{JJ}^{(d)} C_J^{(c)}(L_1S_1) C_J^{(c)}(L_2S_2)
    \nonumber \\
    &\ \ \ 
    +\frac{e^2 \lambdabar^2}{b^3}  \qty[
    - 2R_{JJ}^{(c)} C_J^{(c)}(L_1S_1) C_J^{(c)}(L_2S_2)
    + \sum_s R^{(b)}_{J-s,J}\qty(
    C^{(c)}_J(L_1S_1) C^{(d)}_J(L_2S_2,s) + C^{(d)}_J(L_1S_1,s) C^{(c)}_J(L_2S_2)
    )
    ]
\end{align}
\end{widetext}
for $\lambda=-1$.
We have defined the integrals
\begin{align}
    R^{(a)}_{L_1L_2}(x) &= \int_0^\infty \diff \kappa \frac{\kappa^2}{\kappa^2+x^2} Q_{L_1}(\kappa) Q_{L_2}(\kappa)
    , \\
    R^{(b)}_{L_1L_2}(x) &= \int_0^\infty \diff \kappa \frac{\kappa^3}{\kappa^2+x^2} Q_{L_1}(\kappa) P_{L_2}(\kappa)
    , \\
    R^{(c)}_{L_1L_2}(x) &= \int_0^\infty \diff \kappa \frac{\kappa^4}{\kappa^2+x^2} P_{L_1}(\kappa) P_{L_2}(\kappa)
    , \\
    R^{(d)}_{L_1L_2}(x) &= \int_0^\infty \diff \kappa \frac{\kappa^2}{\kappa^2+x^2} P_{L_1}(\kappa) P_{L_2}(\kappa)
    ,
\end{align}
which are numerically evaluated using atomic wave function in Eq.~\eqref{eq:atomic_radial}.
A $x$-dependence is needed for retardation of the interactions, but in the non-relativistic limit with $c\to \infty$ we only need to use the value at $x=0$: $R^{(a,b,c,d)}_{L_1L_2} \equiv R^{(a,b,c,d)}_{L_1L_2} (x=0)$.

\section{Relation to conventional Slater-Condon parameters}

For comparison, the conventional Coulomb interaction part, $F^{J(\lambda)}_{L_1S_1,L_2S_2}$, are also shown in Tab.~\ref{tab:valF}.
These parameters are related to the Slater-Condon parameter $\mathcal F^{0,2,4,6}$:
\begin{align}
    F^{J(-1)}_{J0,J0} &= H^J \mathcal F^J
    ,
\end{align}
where the constants $H^J$ are listed in the supplementary material of Ref.~\cite{Iimura21}: $H^0=1$, $H^2=\frac{2}{35}$, and $H^4=\frac{2}{63}$ for $l=2$; $H^0=1$, $H^2=\frac{4}{75}$, $H^4=\frac{2}{99}$, and $H^6 = \frac{100}{5577}$ for $l=3$.
The Hund's coupling $J_{\rm H}$ is then determined by the expression in Ref.~\cite{Marel88}:
\begin{align}
    J_{\rm H} &= \frac{\mathcal F^2 + \mathcal F^4}{14}
\end{align}
for $d$ electrons ($l=2$) and 
\begin{align}
    J_{\rm H} &= \frac{286\mathcal F^2 + 195\mathcal F^4 + 250 \mathcal F^6}{6435}
\end{align}
for $f$ electrons ($l=3$).

\begin{table}[b]
\begin{tabular}{cc|cccc||ccccc}
\hline
$J$ & $\lambda$ &  $L_1$ &  $S_1$ &  $L_2$ &  $S_2$ &  $3d$&  $4d$&  $5d$ & $4f$ & $5f$
\\
\hline
0 & $-1$ & 0 & 0 & 0 & 0 &   $16.053$ & $16.434$ & $16.223$ & $21.275$ & $22.714$
\\
2 & $-1$ & 2 & 0 & 2 & 0 &   $0.4842$ & $0.4747$ & $0.4715$ & $0.6357$ & $0.6218$
\\
4 & $-1$ & 4 & 0 & 4 & 0 &   $0.1754$ & $0.1807$ & $0.1825$ & $0.1609$ & $0.1621$
\\
6 & $-1$ & 6 & 0 & 6 & 0 &   ----- & ----- & ----- & $0.1057$ & $0.1102$
\\
\hline
\end{tabular}
 \caption{
Values of the bare conventional Coulomb interaction part, $F^{J(\lambda)}_{L_1S_1,L_2S_2}$, where no screening effects are considered.
The parameters is normalized by the Hund's coupling $J_{\rm H}$.
}
\label{tab:valF}
\end{table}

\vspace{10mm}
\noindent
{\bf \large References}
\\[1mm]
See the list of references in the main text.


\begin{thebibliography}{99}



\bibitem{Fisher89} 
R.A. Fisher, S. Kim, B.F. Woodfield, N.E. Phillips, L. Taillefer, K. Hasselbach, J. Flouquet, A.L. Giorgi, and J.L. Smith, Phys. Rev. Lett. {\bf 62}, 1411 (1989).

\bibitem{Machida12} 
Y. Machida, A. Itoh, Y. So, K. Izawa, Y. Haga, E. Yamamoto, N. Kimura, Y. Onuki, Y. Tsutsumi, and K. Machida
Phys. Rev. Lett. {\bf 108}, 157002 (2012).

\bibitem{Ott84}
H.R. Ott, H. Rudigier, T.M. Rice, K. Ueda, Z. Fisk, and J.L. Smith, Phys. Rev. Lett. {\bf 52}, 1915 (1984).

\bibitem{Shimizu19}
Y. Shimizu, D. Braithwaite, D. Aoki, B. Salce, and J.-P. Brison,
Phys. Rev. Lett. {\bf 122}, 067001 (2019).

\bibitem{Saxena00}
S.S. Saxena, P. Agarwal, K. Ahilan, F. M. Grosche, R. K. W.
Haselwimmer, M. J. Steiner, E. Pugh, I. R. Walker, S. R. Julian, P.
Monthoux, G. G. Lonzarich, A. Huxley, I. Sheikin, D. Braithwaite,
and J. Flouquet, Nature {\bf 406}, 587 (2000).

\bibitem{Aoki01}
D. Aoki, A. Huxley, E. Ressouche, D. Braithwaite, J. Flouquet, J.-P.
Brison, E. Lhotel, and C. Paulsen, Nature {\bf 413}, 613 (2001).

\bibitem{Huy07}
 N. T. Huy, A. Gasparini, D. E. de Nijs, Y. Huang, J. C. P. Klaasse, T.
Gortenmulder, A. de Visser, A. Hamann, T. G\"{o}rlach, and H. v.
L\"{o}hneysen: Phys. Rev. Lett. {\bf 99}, 067006 (2007).

\bibitem{Ran19}
S. Ran, C. Eckberg, Q.-P. Ding, Y. Furukawa, T. Metz,
S.R. Saha, I-L. Liu, M. Zic, H. Kim, J. Paglione, N.P. Butch, Science {\bf 365}, 684 (2019).

\bibitem{Aoki19}
D. Aoki, A. Nakamura, F. Honda, D. Li, Y. Homma, Y. Shimizu,
Y.J. Sato, G. Knebel, J.-P. Brison, A. Pourret, D. Braithwaite,
G. Lapertot, Q. Niu, M. Vali$\check{\rm c}$ka, H. Harima, and J. Flouquet, J. Phys. Soc. Jpn. {\bf 88}, 043702 (2019).

\bibitem{Aoki_review}
For a review, see D. Aoki and J. Flouquet,
J. Phys. Soc. Jpn. {\bf 81}, 011003 (2012);
D. Aoki, J.-P. Brison, J. Flouquet, K. Ishida, G. Knebel, Y. Tokunaga, and Y. Yanase,
J. Phys.: Condens. Matter {\bf 34}, 243002 (2022).

\bibitem{Mydosh11} For a review, see J. A. Mydosh and P. M. Oppeneer: Rev. Mod. Phys. {\bf 83} (2011) 1301.

\bibitem{Bethe_book} H.A. Bethe and E.E. Salpeter, {\it Quantum Mechanics of One- and Two-Electron Atoms} (Dover, New York, 1977).

\bibitem{Itoh65} T. Itoh, Rev. Mod. Phys. {\bf 37}, 159 (1965).

\bibitem{Liu20} For example, see W. Liu, J. Chem. Phys. {\bf 152}, 180901 (2020).

\bibitem{Naito20} T. Naito, R. Akashi, H. Liang, and S. Tsuneyuki, J. Phys. B: At. Mol. Opt. Phys. {\bf 53}, 215002 (2020).

\bibitem{Condon_book} E.U. Condon and G. H. Shortley, {\it The Theory of Atomic Spectra}
(Cambridge University Press, Cambridge, 1951).

\bibitem{Slater_book}
J.C. Slater, Quantum Theory of Atomic Structure (McGraw-
Hill, New York, 1960).

\bibitem{Kanamori63}
J. Kanamori, Prog. Theor. Phys. {\bf 30}, 275 (1963).

\bibitem{Czyzyk94} M. T. Czyzyk and G. A. Sawatzky, Phys. Rev. B {\bf 49}, 14211
(1994).

\bibitem{Aichhorn09} 
M. Aichhorn, L. Pourovskii, V. Vildosola, M. Ferrero, O. Parcollet, T. Miyake, A. Georges, and S. Biermann, Phys. Rev. B {\bf 80}, 085101 (2009).

\bibitem{Iimura21} S. Iimura, M. Hirayama, and S. Hoshino, Phys. Rev. B {\bf 104}, L081108 (2021).

\bibitem{Bjorken_book} J.D. Bjorken and S.D. Drell, {\it Relativistic Quantum Fields} (McGraw-Hill, New York, 1965).

\bibitem{Sucher80} J. Sucher, Phys. Rev. A {\bf 22}, 348 (1980).

\bibitem{Berestetskii_book} V.B. Berestetskii, E.M. Lifshitz, and L.P. Pitaevskii, {\it Quantum Electrodynamics} (Butterworth-Heinemann, Oxford, 1982).

\bibitem{Wang06} Y. Wang, K. Xia, Z.-B. Su, and Z. Ma, Phys. Rev. Lett. {\bf 96}, 066601 (2006).

\bibitem{Hoshino23} S. Hoshino, M.-T. Suzuki, and H. Ikeda, Phys. Rev. Lett. {\bf 130}, 256801 (2023).

\bibitem{suppl} See Supplementary Material which includes citation of Refs.~\cite{Wang17,Varshalovich_book,Itoh65,Marel88,Rose_book} and provides simple derivation of Coulomb-Breit interaction (S1), definitions of atomic wave function (S2), angular momentum tensor (S3), sphreical coordinates (S4), and derivation of generalized Slater-Condon parameters (S5). Conventional Slater-Condon parameter is also listed for reference (S6).

\bibitem{Khveshchenko93} D.V. Khveshchenko, Phys. Rev. B {\bf 47}, 3446 (1993).

\bibitem{Lee07} S.-S. Lee, P.A. Lee, and T. Senthil, Phys. Rev. Lett. {\bf 98}, 067006 (2007).

\bibitem{Lee14} P.A. Lee, Phys. Rev. X {\bf 4}, 031017 (2014).

\bibitem{Schlawin19}
F. Schlawin, A. Cavalleri, and D. Jaksch,
Phys. Rev. Lett. {\bf 122}, 133602 (2019).

\bibitem{Gao20}
H. Gao, F. Schlawin, M. Buzzi, A. Cavalleri, and D. Jaksch,
Phys. Rev. Lett. 125, 053602 (2020).

\bibitem{Chakraborty21}
 A. Chakraborty and F. Piazza, Phys. Rev. Lett. {\bf 127}, 177002 (2021).

\bibitem{Schlawin22}
F. Schlawin, D. M. Kennes, M. A. Sentef, 
Appl. Phys. Rev. {\bf 9}, 011312 (2022).

\bibitem{Bloch22} 
J. Bloch, A. Cavalleri, V. Galitski, M. Hafezi, and A. Rubio, Nature {\bf 606}, 41 (2022).

\bibitem{Negele_book} J. W. Negele and H. Orlando, Quantum Many-Particle Systems (Westview Press, Colorado, 1998).

\bibitem{Nagaosa_book} N. Nagaosa, {\it Quantum Field Theory in Condensed Matter Physics} (Springer, Berlin, 1999).

\bibitem{Gorceix88} O. Gorceix and P. Indelicato, Phys. Rev. A {\bf 37}, 1087 (1988).

\bibitem{Messiah_book} A. Messiah, {\it Quantum Mechanics} (Dover, New York, 1999).

\bibitem{Rose_book} M.E. Rose, {\it Elementary theory of angular momentum} (Dover, New York, 1995).

\bibitem{Varshalovich_book} D.A. Varshalovich, A.N. Moskalev, and V.K. Khersonskii, {\it Quantum Theory of Angular Momentum} (World Scientific, Singapore, 1988).

\bibitem{Sugano_book}
S. Sugano, Y. Tanabe, and H. Kamimura, Multiplets of
Transition-Metal Ions in Crystals (Academic Press, New York,
1970).

\bibitem{Coury16}
M.E.A. Coury, S.L. Dudarev, W.M.C. Foulkes, A.P.
Horsfield, P.-W. Ma, and J. S. Spencer, Phys. Rev. B {\bf 93}, 075101
(2016).

\bibitem{Bunemann17}
J. B\"{u}nemann and F. Gebhard, J. Phys.: Condens. Matter {\bf 29},
165601 (2017).

\bibitem{Letouze23} C. Letouz\'{e}, G. Radtke, B. Lenz, C. Brouder, arXiv:2306.10186 (2023).

\bibitem{Wang17} Y. Wang, H. Weng, L. Fu, and X. Dai, Phys. Rev. Lett. {\bf 119}, 187203 (2017).

\bibitem{Ohkawa83}
F. J. Ohkawa, J. Phys. Soc. Jpn. {\bf 52}, 3897 (1983).

\bibitem{Shiina97}
R. Shiina, H. Shiba, and P. Thalmeier, J. Phys. Soc. Jpn. {\bf 66},
1741 (1997).

\bibitem{Kuramoto00}
Y. Kuramoto and H. Kusunose, J. Phys. Soc. Jpn. {\bf 69}, 671
(2000).

\bibitem{Santini00}
P. Santini and G. Amoretti, Phys. Rev. Lett. {\bf 85}, 2188 (2000).

\bibitem{Kiss05}
A. Kiss and Y. Kuramoto, J. Phys. Soc. Jpn. {\bf 74}, 2530
(2005).

\bibitem{Takimoto05}
T. Takimoto, J. Phys. Soc. Jpn. {\bf 75}, 034714 (2005).

\bibitem{Kusunose08}
H. Kusunose, J. Phys. Soc. Jpn. {\bf 77}, 064710 (2008).

\bibitem{Kuramoto09}
For a review, see, Y. Kuramoto, H. Kusunose, and A. Kiss,
J. Phys. Soc. Jpn. {\bf 78}, 072001 (2009).

\bibitem{Haule09}
K. Haule and G. Kotliar, Nat. Phys. {\bf 5}, 796 (2009).

\bibitem{Ikeda12}
H. Ikeda, M.-T. Suzuki, R. Arita, T. Takimoto, T. Shibauchi,
and Y. Matsuda, Nat. Phys. {\bf 8}, 528 (2012).

\bibitem{Suzuki17}
M.-T. Suzuki, T. Koretsune, M. Ochi, and R. Arita, Phys.
Rev. B {\bf 95}, 094406 (2017).

\bibitem{Hayami18}
S. Hayami, M. Yatsushiro, Y. Yanagi, and H. Kusunose,
Phys. Rev. B {\bf 98}, 165110 (2018).

\bibitem{Chikano21} N. Chikano, S. Hoshino, and H. Shinaoka, Phys. Rev. B {\bf 104}, 235125 (2021).

\bibitem{Kusunose20} H. Kusunose, R. Oiwa, and S. Hayami, J. Phys. Soc. Jpn. {\bf 89}, 104704 (2020).

\bibitem{Marel88} D. van der Marel and G.A. Sawatzky, Phys. Rev. B {\bf 37}, 10674 (1988).

\bibitem{AtomicWebSite} A. Kramida, Yu. Ralchenko, J. Reader, and NIST ASD Team (2022). NIST Atomic Spectra Database (ver. 5.10), [Online]. Available: https://physics.nist.gov/asd [2023, November 8]. National Institute of Standards and Technology, Gaithersburg, MD.

\bibitem{Clementi67} E. Clementi, D.L. Raimondi, and W.P. Reinhardt, J. Chem. Phys. {\bf 47}, 1300 (1967).













\end{thebibliography}
\end{document}